\documentclass[aps,prb,twocolumn,superscriptaddress,nobibnotes,10pt]{revtex4-1}
\usepackage[utf8]{inputenc}
\usepackage[T1]{fontenc}
\usepackage{lmodern}
\usepackage{hyperref}
\usepackage[hyphenbreaks]{breakurl}
\usepackage{tikz}
\usetikzlibrary{arrows,petri}
\usepackage{verbatim}
\usepackage{xstring}
\usepackage{multirow,tabularx}
\usepackage{colortbl}
\usepackage{colordvi}

\newcommand{\aiida}{\emph{AiiDA}}
\newcommand{\aiidav}{v0.8.1}
\newcommand{\qe}{\emph{Quantum ESPRESSO}}

\newcommand{\ha}{$E_h$}
\newcommand{\bohr}{$a_0$}

\newcommand{\tcodid}[1]{\href{http://www.crystallography.net/tcod/#1.html}{#1}}

\def\subst#1#2#3{%
  \IfSubStr{#1}{#2}{%
      \StrSubstitute{#1}{#2}{#3}}{#1}}

\newcommand{\tag}[1]{\texttt{\subst{#1}{_}{\_}}}
\newcommand{\dic}[1]{\texttt{\subst{#1}{_}{\_}.dic}}
\newcommand{\script}[1]{\texttt{\subst{#1}{_}{\_}}}

\newcommand\THEOSMARVEL{Theory and Simulation of Materials (THEOS) and National 
                        Center for Computational Design and Discovery of Novel Materials (MARVEL),
                        {\'E}cole Polytechnique F{\'e}d{\'e}rale de Lausanne, 1015, Switzerland}

\newcommand\VILNIUS{Vilnius University Institute of Biotechnology,
                    Saul{\.{e}}tekio al.\ 7, LT-10257 Vilnius, Lithuania}

\newcommand\VUMIF{Vilnius University Faculty of Mathematics and Informatics,
                  Naugarduko st.\ 24, LT-03225, Vilnius, Lithuania}

\begin{document}

\title{A posteriori metadata from automated provenance tracking:
       Integration of AiiDA and TCOD}

\author{Andrius Merkys}
\affiliation{\THEOSMARVEL}
\affiliation{\VILNIUS}
\author{Nicolas Mounet}
\affiliation{\THEOSMARVEL}
\author{Andrea Cepellotti}
\affiliation{\THEOSMARVEL}
\author{Nicola Marzari}
\affiliation{\THEOSMARVEL}
\author{Saulius~Gra\v{z}ulis}
\affiliation{\VILNIUS}
\affiliation{\VUMIF}
\author{Giovanni Pizzi}
\affiliation{\THEOSMARVEL}

\begin{abstract}
In order to make results of computational scientific research 
findable, accessible, interoperable and re-usable, it is necessary 
to decorate them with standardised metadata. However, there are a 
number of technical and practical challenges that make this process 
difficult to achieve in practice. Here the implementation of a 
protocol is presented to tag crystal structures with their computed 
properties, without the need of human intervention to curate the 
data. This protocol leverages the capabilities of \aiida{}, an 
open-source platform to manage and automate scientific computational 
workflows, and TCOD, an open-access database storing computed 
materials properties using a well-defined and exhaustive ontology. 
Based on these, the complete procedure to deposit computed data in 
the TCOD database is automated. All relevant metadata are extracted 
from the full provenance information that \aiida{} tracks and stores 
automatically while managing the calculations. Such a protocol also 
enables reproducibility of scientific data in the field of 
computational materials science. As a proof of concept, the 
\aiida{}-TCOD interface is used to deposit 170 theoretical structures 
together with their computed properties and their full provenance 
graphs, consisting in over 4600 \aiida{} nodes.

\end{abstract}

\maketitle

\section{Introduction}
Modelling and simulation are commonly identified as the third 
paradigm in scientific understanding, complementing theory and 
experiments. In particular, computational materials science has 
developed into an essential field due to two main factors. First, in 
the past years significant advances have been achieved both in the 
approximations of the theories to simulate materials from 
first-principles~\cite{Jain2016} and in the codes that implement 
them (many of which are distributed open-source). Second, these 
computationally-expensive calculations have been made feasible 
thanks to the exponential increase of computing power predicted by 
Moore's law and the corresponding decrease of the price/performance 
ratio. As a consequence, large number of properties can nowadays be 
computed for large families of materials. A number of online 
databases has appeared in the past few years, like the Materials 
Project~\cite{Jain2013}, OQMD~\cite{Saal2013} and AFLOWLIB~\cite
{Curtarolo2012}. However, much effort is still needed to consolidate 
the knowledge from publications, tagging results with suitable 
metadata under an established ontology, and preserving at the same 
time the complete provenance of the computed data to enable 
reproducibility of the results.

Currently, there are several attempts to define an ontology in the
field of theoretical material science, like the European Theoretical
Spectroscopy Facility (ETSF)~\cite{Caliste2008,Gonze2008a},
NoMaD~\cite{NoMaDMetaInfo2016}, OPTiMaDe~\cite{OPTiMaDe2017}
and the Theoretical Crystallography Open Database
(TCOD)~\cite{TCODWebPage2016,Grazulis2014}. The latter
has been launched with
the aim to collect the results of several kinds of calculations (DFT,
post-HF, QM/MM, etc.), into an open-access resource for long-term
archival storage. TCOD adopts the Crystallographic Information
Framework~(CIF) format~\cite{Hall1991}, a unified format for
reporting and storing the results of experimentally-solved crystal
structures, which has been widely adopted and used as the
\emph{de facto} standard by most crystallographic journals as well as
structural databases like, to mention just a few, the Inorganic
Crystal Structure Database~(ICSD)~\cite{Belsky2002}, the Cambridge
Structural Database~\cite{Groom2014}, the American Mineralogist
Crystal Structure Database~\cite{Rajan2006}, and the Crystallography
Open Database (COD)~\cite{Grazulis2017,Grazulis2012}. One of the main
advantages of the CIF format is the existence of CIF dictionaries,
aimed at defining domain-specific ontologies readable both by humans
and by machines~\cite{Brown2002}. Constraints, units of measurement
and interrelationships for data values are specified in order to
homogenise the data, eliminate ambiguities and allow for automated
validation. Furthermore, since CIF data names are uniquely defined, a
CIF file may contain properties from more than one dictionary, making
it possible to easily extend and complement the file (e.g. for
macromolecular crystallography~\cite{Fitzgerald2006a}, powder
diffraction~\cite{Toby2003}, electron density~\cite{Mallinson2005}
and experimental material properties~\cite{Pepponi201210}).
Currently, CIF format is developed to its 2.0 version
(CIF~2,~\cite{Bernstein2016}) with even more features for ontology
definition. In TCOD domain-specific dictionaries have been compiled in
order to define an ontology for the hosted data (in particular, the
\dic{cif\_dft} dictionary for DFT-based properties, and \dic{cif\_tcod}
for the generic metadata related to scientific workflows), and
automated checks of CIF files against these dictionaries have been
implemented.

The definition of an ontology is not the only challenge that 
materials science is facing; another major issue is the preservation 
of provenance for result replication. In fact, currently most of the 
scientific publications provide only a subset of all control 
parameters, numerical inputs and calculation interrelationships 
needed to exactly reproduce the published results. This problem can 
be solved by using provenance-tracking frameworks like \aiida{}~
\cite{AiiDA2017,Pizzi2016}, a high-throughput infrastructure that 
provides a high-level research environment to automate the execution 
of computations, systematically store inputs and outputs and their 
relationships in a graph database (tailored to keep track of the 
full data provenance) and share results.

In this work we present the integration of the TCOD database with 
\aiida{}, using and enhancing the \dic{cif\_tcod} CIF
dictionary~\cite{TCODMailingList2016}. Our integration of a
calculation automation framework and an ontology-based database
allows for the \emph{a posteriori} deposition of simulation results
with automatically recorded metadata. Most importantly, the metadata 
tagging and deposition can be performed at any time after the 
calculation has been executed thanks to the automatic provenance 
tracking provided by \aiida{}.

In the following, we first describe the provenance model implemented 
in \aiida{} and explain how we map it to a CIF file. We explain how 
we address and solve the technical issues that arise in the process, 
e.g. the inclusion of input and output files within the CIF file and 
their encoding, how software versions can be tracked, and how to 
report bibliographic references. We then describe the extensions to 
the TCOD dictionary that we implemented, and compare the latter with 
other existing ontologies. The algorithm of the converters to 
integrate \aiida{} and TCOD is then illustrated. Finally, we discuss 
the results obtained and deposited in TCOD using the codes and 
algorithms implemented in this work.

\section{Workflow representation}
\subsection{Provenance model and directed acyclic graphs}
\label{sec:provenancemodel}

\begin{figure*}[tbp]
  \caption{Workflow of TCOD entry~\tcodid{10000008}.
           Workflow consists of three
           consecutive relaxations of GaGeO$_3$ structure with the \qe{}
           pw.x code. Artifacts (called ``data nodes'' in \aiida{}) are
           marked as circles and processes (called ``calculation nodes''
           in \aiida{}) as rectangles. A special type of \aiida{} artifact,
           a code, is represented by a diamond. Note that in \aiida{} the
           direction of arrows is inverted with respect to the OPM
           notation~\cite{Moreau2007,Moreau2008}.}

  \label{2048-graph}
  \input{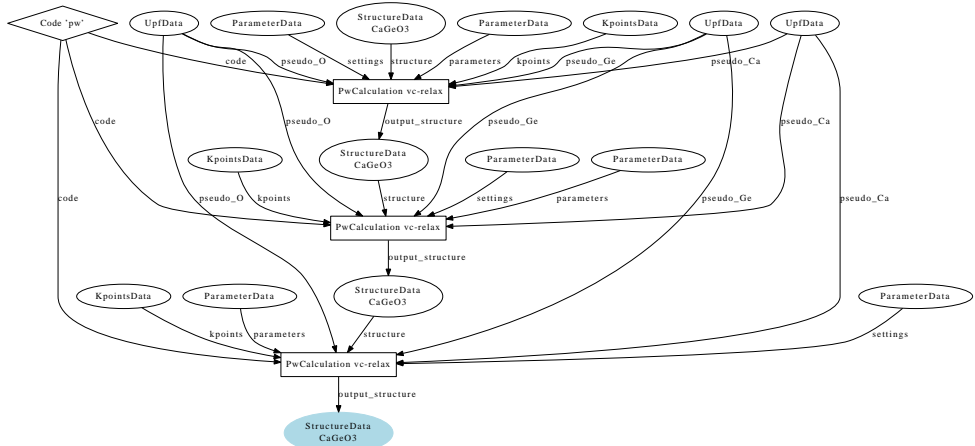}
\end{figure*}

Reproducibility of scientific calculations is a crucial tenet of
computational scientific research~\cite{Mesirov2010,Peng2006,Peng2009},
and it would be tremendously facilitated by a system enabling easy
replication of exact or modified computations reported in a scientific
publication~\cite{Peng2011}. An important prerequisite is that content
(data) must be separated from its presentation (article)~\cite{Peng2008}.
Such separation is implemented for instance in
\emph{Sweave}~\cite{Leisch2008}, designed specifically for the \emph{R}
statistical programming language~\cite{RCoreTeam2016}. In the field of
Computational Materials Science, Pizzi \emph{et~al}.\ have developed
\aiida{}, a \emph{Python}-based framework for atomistic
simulations~\cite{Pizzi2016}, where data provenance is stored
automatically while the simulations and workflows are executed, using a
data model elaborating on ideas of the Open Provenance
Model~(OPM)~\cite{Moreau2007,Moreau2008}. OPM suggests to represent whole
scientific workflows of data transformations as directed acyclic
graphs~(DAGs), whose nodes are \emph{artifacts} and vertices are
\emph{processes}. An example of DAG of \aiida{} workflow is shown in
Fig.~\ref{2048-graph}.

The purpose of this work is to represent faithfully the full 
provenance of computational workflows, represented by \aiida{} in 
the form of DAGs, within in a CIF file. To achieve this goal, we 
have applied \dic{cif_tcod} CIF dictionary as follows. We represent 
workflows by an ordered list of processes (``workflow steps''), so 
that the execution of such sequence leads to the generation of the 
workflow results. Workflow steps are represented in a CIF loop using 
\tag{_tcod_computation_*} data items, and the sequential numbers of 
workflow steps are given in \tag{_tcod_computation_step}. Each step 
is then defined by its command line string (\tag
{_tcod_computation_command}) and the environment variables (\tag
{_tcod_computation_environment}).

Input and output files and directories (artifacts in the OPM) are 
described in a CIF loop of \tag{_tcod_file_*} data items. For the 
sake of achieving a deterministic order, we recommend to provide 
files and directories sorted by lexicographical order of their full 
path, with directories always preceding their contents. Moreover, 
all paths should be relative to the same location throughout all 
calculations. In particular, the data item \tag{_tcod_file_name} is 
used for names of files and directories, whereas the file content is 
stored in the \tag{_tcod_file_contents} data item (to comply with 
the requirements of the CIF file format, the file contents are 
encoded as described in Sec.~\ref{sub:encodings}). Since directories 
do not have file-like contents, the special CIF value `.' (a dot, 
meaning ``data are inapplicable'') must be provided for them. 
Contents of standard input, output and error of processes (if any) are
placed in separate files and linked to the workflow steps using \tag
{_tcod_computation_input_file}, \tag {_tcod_computation_stdout} and 
\tag{_tcod_computation_stderr} data items, accordingly.

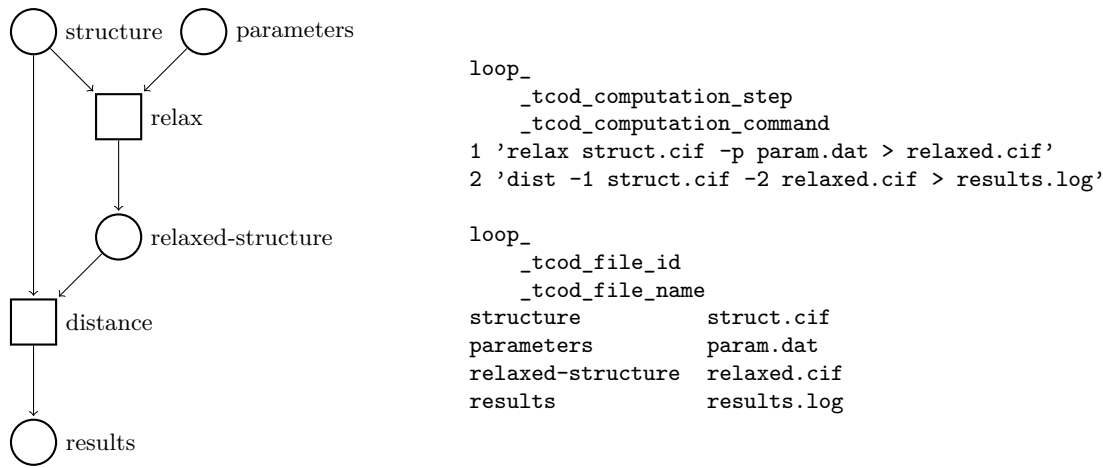
\begin{figure*}[tbp]
  \caption{Example of a simple workflow~\cite{Pizzi2016}.
           Graph view~(\emph{left}) and representation in TCOD
           CIF~(\emph{right}). In the graph view, files are presented
           as circles and processes as squares. For the sake of brevity
           and clarity, file contents are not reported here for the
           TCOD CIF representation.}

  \label{wf-example}

  \begin{minipage}{0.42\textwidth}
    \begin{tikzpicture}[node distance=1.6cm]
      \tikzstyle{file}=[circle,thick,draw=black,fill=white,minimum size=6mm]
      \tikzstyle{process}=[rectangle,thick,draw=black,fill=white,minimum size=6mm]

      \node[file,label=right:structure] (f1) {};
      \node[process,label=right:relax] (p1) [below right of=f1] {}
        edge [pre] (f1);
      \node[file,label=right:parameters] (f2) [above right of=p1] {}
        edge [post] (p1);
      \node[file,label=right:relaxed-structure] (f3) [below of=p1] {}
        edge [pre] (p1);
      \node[process,label=right:distance] (p2) [below left of=f3] {}
        edge [pre] (f1)
        edge [pre] (f3);
      \node[file,label=right:results] (f4) [below of=p2] {}
        edge [pre] (p2);
    \end{tikzpicture}
  \end{minipage}%
  \begin{minipage}{0.58\textwidth}
    \small
\begin{verbatim}
loop_
    _tcod_computation_step
    _tcod_computation_command
1 'relax struct.cif -p param.dat > relaxed.cif'
2 'dist -1 struct.cif -2 relaxed.cif > results.log'

loop_
    _tcod_file_id
    _tcod_file_name
structure          struct.cif
parameters         param.dat
relaxed-structure  relaxed.cif
results            results.log
\end{verbatim}
  \end{minipage}
\end{figure*}

While defining this format, we needed to address a technical issue. The
CIF format poses restrictions on acceptable CIF values, so that there are
cases in which the contents of a file must be encoded (the most obvious
case is a file containing a single `.' symbol, that would otherwise be
ignored). We have thus developed a few content-encoding protocols that we
describe in detail in Section~\ref{sub:encodings}, allowing inclusion of
any text and binary file content into CIF text fields. An example of a
simple workflow along with its representation in the TCOD CIF format is
given in Fig.~\ref{wf-example}.

\subsection{Input data}
\label{sec:inputdata}

Initial steps of atomistic simulation workflows usually transform input
data (often from external sources or databases) to internal structures. To
preserve the full history, it is crucial to maintain a reference to the
original data. This is relatively straightforward if the resource is
available on the Internet \emph{and} is assigned a permanent URI, DOI
or similar resource identifier. Usually, upon retrieval, it is beneficial
to supplement such identifier with retrieval date, resource version (if
given) and a checksum as a tool for ultimate integrity control.

As part of this work, we have implemented a number of external database
importers as part of \aiida{}. These provide the user with the possibility
to seamlessly import data from several structural databases, including
COD~\cite{Grazulis2012}, ICSD~\cite{Belsky2002}, Material Properties Open
Database~(MPOD)~\cite{Pepponi201210}, Open Quantum Materials
Database~(OQMD)~\cite{Kirklin2015}, as well as a pseudopotential database,
the NNIN/C Pseudopotential Virtual Vault~\cite{NNINC2016}. By means of
these importers, the user can both query and fetch data from the
respective databases, and then import the relevant entries directly into
\aiida{}. During this process, we make sure to record information about
the source: permanent URIs (when available), versions and checksums. When
the \aiida{} graph is later exported to CIF format and deposited in TCOD,
the source of the initial structural data is recorded in the
\dic{cif_tcod} data items \tag{_tcod_source_structure_*} and
\tag{_tcod_source_database_*}.

\subsection{File content inclusion in CIF and encoding}
\label{sub:encodings}

As we described in Sec.~\ref{sec:provenancemodel}, in TCOD CIFs  we aim at
storing an archive of the whole computational workflow used to obtain the
properties of a given material within the same TCOD CIF file. As a
consequence, the file contents of individual calculation inputs and
outputs are stored as values of the \tag{_tcod_file_contents} data item.
However, due to the format restrictions of the CIF format, not all data
can be stored unmodified in a CIF~1.1 file text field. In particular, the
following restrictions apply:

\begin{itemize}
  \item the character set is restricted to printable and whitespace
        ASCII characters, so that CIF files must not contain unescaped
        binary data and Unicode symbols;
  \item lines (except the first) must not start with semicolons. This
        character is reserved as the text field delimiter. Such
        limitation forbids nesting CIF-inside-a-CIF, but is effectively
        averted using line prefixing protocol~\cite{Bollinger2011},
        non-standard in CIF~1.1, albeit included in the recent CIF~2
        specification~\cite{Bernstein2016};
  \item line lengths typically have to follow a number of
        recommendations~\cite{Merkys2016}.
\end{itemize}

Since no standard solution exists for all mentioned issues of 
arbitrary data presentation for CIF~1.1, we have devised and 
implemented a protocol to encode and decode file contents prior to 
their storage in CIF text fields. We implemented a few encoding 
schemes because of their different, non-overlapping advantages:

\begin{itemize}
  \item {\bf Numeric Character Reference} (NCR): used in \emph{cod-tools}
    package~\cite{Merkys2016} to escape binary content and semicolons at
    the start of lines. This method retains the readability of the text
    with sparse non-ASCII symbols;
  \item {\bf Quoted-Printable} \cite{Freed1996}: same properties as NCR
    with in addition the ability to fold long lines;
  \item {\bf Base64}~\cite{Freed1996}: overcomes all the deficiencies
    of CIF at the price of readability and file size; chosen only when
    the file content is purely binary;
  \item {\bf gzip+Base64}~\cite{Freed1996,Deutsch1996}: same as Base64
    with additional file compression.
\end{itemize}

As one may notice, the gzip+Base64 encoding is composite: it defines a
stack of two encodings -- Base64-encoding of gzip'ed contents. To
accommodate this and possibly other composite encodings, we have defined a
set of \tag{_tcod_content_encoding_*} data items, allowing to describe any
complex stack of encodings in a CIF loop.

The choice of the encoding is arbitrary and dependent on the requirements
of readability and file size expected by the user. Nevertheless, to be
able to automate the process of TCOD CIF file generation, we have
implemented an algorithm to automatically choose the most appropriate
encoding while trying to preserve maximum data readability, as described
in Sec.~\ref{sec:implementation}.

To ensure the integrity of both plain and encoded files, checksums are
recorded alongside file contents. As of \aiida{}~\aiidav{}, both MD5
and SHA1 algorithms are used.

\subsection{Software versioning}

While problems of input data versioning may be avoided with revision
control systems, WORM (Write Once Read Many) databases, permanent links
and checksums, keeping proper description of data transformations remains
a challenging task. In particular, the algorithm of each transformation
should be specified in a strictly-defined, machine-readable form. In
addition, it would be extremely useful, if not essential, to be able to
assert if two different representations will provide the same output when
the same input is provided. Basic informatics principles, however, pose a
limit on the applicability of such description, since in general there
exists no universal algorithm that can establish the equivalence of two
Turing-complete language programs by formally analysing
them~\cite{Turing1937,Rice1953}. In practice, however, algorithms in
normalised form could be recorded and claimed to perform identical
transformations if their normalised forms are identical. This
is hard to achieve, however, since these ``descriptions'' must be expanded
to include compilation/interpretation tools and environment, runtime
operating system (OS) and external dynamically-linked libraries. Ideally
these parameters should be collected recursively for every dependency.
The CPU introduces a final caveat, since two different CPUs, in particular if one or both
of them are buggy, can interpret the same algorithm in different ways yielding
different results~\cite{Moler1995}. Thus, in addition to provenance,
``descriptions'' may be indispensable for an efficient bug tracking.
As of \aiida{}~\aiidav{}, data transformations are described by the internal
location of executables, runtime command line parameters, environment,
execution time stamp and scheduler directives. Moreover, codes from
atomistic simulation packages, such as \qe{}~\cite{Giannozzi2009} and
\emph{NWChem}~\cite{Valiev2010}, which are interfaced with \aiida{}, can
be queried for version, compilation and runtime parameters. In addition to
this, manipulations in the native \emph{Python} environment (referred in
\aiida{} as ``workfunctions'', or ``inline calculations'' in earlier
\aiida{} versions), are also supplemented with the source code. In our
experience, this representation is sufficient at the moment and can be
easily extended in the future, for instance using virtual machines
(e.g. QEMU~\cite{QEMU2017}, VirtualBox~\cite{VirtualBox2017},
VMware~\cite{VMware2017} or similar techniques), or
Docker~\cite{Docker2016}, that has rapidly become a widespread tool to
reproduce a given computational environment. The presence of full
process provenance information in machine-readable format would enable
to perform such reconstructions automatically.

\subsection{Bibliographic references}

Bibliographic references often appear in CIF files as identifiers of
applied algorithms and parameters. A mechanism to provide references in a
machine-readable way is described in the \dic{cif_core} dictionary.
According to the recommendations of the International Union of
Crystallography (IUCr), citation details should be given in structured
tables (CIF loops). In order to categorize references according to the
described aspect of the computation (force field, software code etc.), we
have introduced a data item \tag{_tcod_citation_linkage} with enumeration
values of \tag{force-field}, \tag{software-code}, \tag{model},
\tag{pseudopotential}, \tag{XC-functional} and \tag{basis-set}, to
facilitate automatic classification and filtering of computational details
based on these attributes. More detailed human-readable description of the
relevance can be supplied using \tag{_citation_special_details} data item.

\section{Ontologies for DFT}
\label{sub:ontologies}

The CIF dictionary \dic{cif_dft}, developed by the advisory board of TCOD,
provides data items for the description of basis sets, pseudopotentials,
atomic settings and exchange-correlation functionals. To accommodate input
parameters and calculation results exported from \aiida{}, we have
supplemented the \dic{cif_dft} dictionary with data items for Brillouin
zone, kinetic energy cut-offs and calculated structure properties, such as
the total energy. We have also added bulk modulus and stiffness
tensor, the latter being represented as a symmetric matrix of 21
independent variables. For the sake of simplicity and consistency with the
common practices in the core CIF dictionary, we have followed the \dic{cif_core}'s
approach to represent matrices in separate ``plain'' data items (for
example, standard anisotropic atomic displacement components in
\dic{cif_core}'s \tag{_atom_site_aniso_U_*} data items) instead of
relational database-style loops, used e.g. by the
MPOD~\cite{Pepponi201210}. Fig.~\ref{20000419} shows a sample from a TCOD
CIF file, containing some newly introduced CIF data items.

\begin{figure}[tbp]
  \caption{Sample from TCOD entry~\tcodid{20000419},
           displaying computational setup and bulk modulus (in~GPa),
           convergence criterion for cell energy and kinetic energy
           cut-off for wavefunctions (both in~eV). Units for each data
           item are unambiguously defined in the TCOD dictionary.}

\begin{verbatim}
_dft_bulk_modulus                5.95
_dft_cell_energy_conv            0.00000001
_dft_BZ_integration_method       Monkhorst-Pack
_dft_kinetic_energy_cutoff_wavefunctions 500
_dft_pseudopotential_type        PAW
_dft_XC_functional_type          GGA
_tcod_database_code              20000419
_tcod_model                      DFT
_tcod_software_package           VASP
\end{verbatim}
  \label{20000419}
\end{figure}

The ontology defined in the \dic{cif_dft} dictionary is supplemented by
the one that we adopt in this work. For completeness, we mention here that
other projects have also invested effort in standardising ontologies for
computational materials science simulations, especially for atomistic
and/or DFT-based methods. In particular, we report a comparison of the
ontology in TCOD CIF with those defined by two other projects, ETSF and
NoMaD, in Table~\ref{tbl:etsf-tcodcif} and Table~\ref{tbl:nomad-tcodcif},
respectively. Differences are mainly in notations and conventions: for
instance, ETSF uses hartree and bohr as main measurement units, NoMaD uses
joule and meter, whereas TCOD uses electronvolt and angstrom. As another
example,  TCOD choice of using lengths and angles of basis vectors stems
from experimental crystallography, while ETSF and NoMaD use vector
notation, more common in theoretical materials science. Nevertheless,
automatic conversion between the two formats would be easy to implement,
making it possible to seamlessly share data between different projects.
The TCOD is actively working in contact with other projects to ensure
the possibility of such conversion.

\begin{table*}[tbp]
  \caption{Comparison of a selection of TCOD CIF data items with respect
           to the corresponding ETSF variables.}
  \centering

    \scriptsize

    \begin{tabularx}{0.95\linewidth}{l l X}
    ETSF variable & TCOD CIF data item(s) & comments \\
    \hline
    \tag{valence_charges} & \tag{_dft_atom_type_valence_electrons} & \\
    \rowcolor[gray]{0.75}
    \tag{pseudopotential_types} & \tag{_dft_pseudopotential_type} & \\
    \tag{basis_set} & \tag{_dft_basisset_type} & \\
    \rowcolor[gray]{0.75}
    \tag{exchange_functional} & \tag{_dft_XC_exchange_functional} &  \\
    \tag{correlation_functional} & \tag{_dft_XC_correlation_functional} &  \\
    \rowcolor[gray]{0.75}
    \tag{fermi_energy} (\ha{}) & \tag{_dft_fermi_energy} (eV) & \\

    \multirow{2}{*}{\tag{smearing_scheme}} & \tag{_dft_BZ_integration_smearing_method} & \multirow{3}{2.6in}{ETSF appends M-P order to the scheme, TCOD CIF has a separate data item} \\
    & \tag{_dft_BZ_integration_MP_order} & \\
    & & \\

    \rowcolor[gray]{0.75}
    \tag{smearing_width} & \tag{_dft_BZ_integration_smearing_width} & \\
    \tag{kinetic_energy_cutoff} (\ha{}) & \tag{_dft_kinetic_energy_cutoff_wavefunctions} & \multirow{2}{2.6in}{in ETSF it is not clear whether the variable applies to wavefunctions or charge densities} \\
    & (eV) & \\
    \rowcolor[gray]{0.75}
    \tag{kpoint_grid_shift} & \tag{_dft_BZ_integration_grid_shift_[XYZ]} &  \\

    \multirow{2}{*}{\tag{primitive_vectors} (\bohr{})} & \tag{_cell_length_[abc]} (\AA) & \\
    & \tag{_cell_angle_[alpha,beta,gamma]} & \\

    \rowcolor[gray]{0.75}
    \tag{reduced_symmetry_matrices} & & \\
    \rowcolor[gray]{0.75}
    \tag{reduced_symmetry_translations} & \multirow{-2}{*}{\tag{_space_group_symop_operation_xyz}} & \multirow{-2}{3in}{ETSF provides matrices, TCOD CIF uses string notation} \\

    \tag{space_group} & \tag{_space_group_IT_number} & CIF has 230 spacegroups, ETSF allows for a range from 1~to~232 \\

    \rowcolor[gray]{0.75}
    \tag{reduced_atom_positions} & \tag{_atom_site_fract_[xyz]} &  \\
    \tag{atom_species} & \tag{_atom_site_type_symbol} & \\

    \rowcolor[gray]{0.75}
    \tag{atom_species_names} & & \\
    \rowcolor[gray]{0.75}
    \tag{atomic_numbers} & & \\
    \rowcolor[gray]{0.75}
    \tag{chemical_symbols} & \multirow{-3}{*}{\tag{_atom_site_type_symbol}} & \\

    \tag{reduced_coordinates_of_kpoints} & \tag{_dft_BZ_integration_grid_IBZ_point_[XYZ]} & \\
    \rowcolor[gray]{0.75}
    \tag{kpoint_weights} & \tag{_dft_BZ_integration_grid_IBZ_point_weight} & \\
    \end{tabularx}
  \label{tbl:etsf-tcodcif}
\end{table*}

\begin{table*}[tbp]
  \caption{Comparison of a selection of TCOD CIF data items with respect
           to the corresponding NoMaD metadata.}
  \centering

    \scriptsize

    \begin{tabularx}{0.95\linewidth}{l l X}
    NoMaD metadata & TCOD CIF data item(s) & comments \\
    \hline

    \tag{atom_label} & \tag{_atom_site_label} & \\
    \rowcolor[gray]{0.75}
    \tag{atom_position} & \tag{_tcod_atom_site_initial_fract_[xyz]} & NoMaD uses Cartesian coordinates, TCOD uses fractional coordinates \\
    \tag{basis_set_plan_wave_cutoff} (J) & \tag{_dft_kinetic_energy_cutoff_wavefunctions} (eV) & \\
    \rowcolor[gray]{0.75}
    \tag{configuration_periodic_dimensions} & \tag{_dft_cell_periodic_BC_[XYZ]} & \\

    \multirow{2}{*}{\tag{simulation_cell}} & \tag{_cell_length_[abc]} & \multirow{2}{*}{NoMaD provides vectors} \\
    & \tag{_cell_angle_[alpha,beta,gamma]} & \\

    \rowcolor[gray]{0.75}
    \tag{program_compilation_datetime} & \tag{_tcod_software_package_compilation_timestamp} & NoMaD in Unix timestamp, TCOD CIF in ISO~8601 \\
    \tag{program_name} & \tag{_tcod_software_package} & \\
    \rowcolor[gray]{0.75}
    \tag{program_version} & \tag{_tcod_software_package_version} & \\
    \tag{atom_forces} (N) & \tag{_tcod_atom_site_resid_force_Cartn_[xyz]} (eV/\AA) & \\
    \rowcolor[gray]{0.75}
    \tag{energy_total} (J/atom) & \tag{_tcod_total_energy} (eV) & \\
    \tag{source_references} & \tag{_tcod_source_*} & NoMaD seems to give a free-text field to identify the source of data \\
    \rowcolor[gray]{0.75}
    \tag{time_calculation} & \tag{_tcod_computation_wallclock_time} & \\

    \end{tabularx}
  \label{tbl:nomad-tcodcif}
\end{table*}

\section{Implementation: exporting the data to TCOD}
\label{sec:implementation}
The main outcome of this work is the definition and implementation of
procedures to export the results of theoretical computations managed with
\aiida{} into CIF files and deposit them into the TCOD database. To
achieve this, we have implemented a converter that, starting from a
user-specified structure within the \aiida{} database, is able to create a
CIF format file. This converter allows for complete automatic
\emph{a posteriori} tagging of structures with their metadata. This 
is made possible by analysing the full provenance 
(stored in the \aiida{} DAG) of the final crystal 
structure, extracting/converting all relevant
information, and storing it in the appropriate CIF fields defined in the TCOD
dictionaries discussed before. The generation of CIF
files has been obtained by interfacing \aiida{} with the \emph{cod-tools}
package~\cite{Merkys2016}. We summarize here the steps of the export and
deposition procedure:

\begin{enumerate}
  \item {\bf Conversion of periodic structure from internal \aiida{}
        representation to CIF}. As of \aiida{}~\aiidav{}, there are two types of
        representations of periodic structures in \aiida{}: structure and
        trajectory. A structure can be straightforwardly represented in
        CIF, whereas separate steps of a trajectory can be converted into
        structures. One or both of these conversions are used to produce
        an initial template CIF file (containing \dic{cif_core} data items
        only), which is supplemented by additional data in the following
        steps.
  \item {\bf Detection of the symmetry and reduction of the unit cell}. In
        \aiida{}, modelled materials are represented as non-reduced unit
        cells of a crystal, in other words, as if their symmetry space
        group were \emph{P1}. Such structures have to be reduced to an
        asymmetric unit (if possible), leaving out the symmetrically equivalent atoms. To
        accomplish it, we have harnessed the algorithm by Grosse-Kunstleve
        and Adams~\cite{Grosse-Kunstleve2002a}, using the implementation
        in \emph{spglib}~\cite{Togo2009}.
  \item {\bf Addition of structure properties (total energies, residual
        forces etc.)}. As much data as possible is parsed from the
        output of a computation and added to the CIF data items defined by
        \dic{cif_tcod} and \dic{cif_dft} dictionaries, including energy
        terms and convergence criteria. We have developed a layer to
        convert the output parameters parsed from the computation outputs
        by \aiida{} into data items of TCOD CIF dictionaries. Currently,
        the conversion is implemented for both the \texttt{pw.x} and
        \texttt{cp.x} codes of \qe{}, as well as for the \emph{NWChem}
        package, but the converter has been designed with a modular
        interface and it can thus be easily extended for any other code
        interfaced with \aiida{}.
  \item {\bf Addition of the metadata for reproduction of the results}.
        Since all metadata required for the reproduction of computations
        (files, scripts, command line strings, etc.) are stored by
        \aiida{}, they are easily collected and stored within the CIF
        file, along with the description of each \aiida{} node used in the
        exported workflow. Files, consisting of more than a quarter
        non-ASCII symbols, are assumed to be binary and encoded with
        Base64, whereas other files with fewer non-ASCII symbols, very
        long lines or other features, that could cause CIF parsing errors,
        are encoded with Quoted-Printable. Files, larger than one kilobyte
        (a default value), are gzip'ed, if requested by the user. There is
        an option to exclude the contents of files that could be
        downloaded from the Web via provided URIs thus reducing the size
        of the resulting CIF file. Checksums are recorded in every case
        to ensure the integrity of files.
  \item {\bf Deposition of generated CIF to TCOD}. The final step is the
        upload and deposition of the CIF file in TCOD using the HTTP protocol 
        implemented by TCOD. The
        deposition is initiated as an \aiida{} calculation, wrapping
        the generic command line script \script{cif_cod_deposit} that is part of the
        \emph{cod-tools} package. \script{cif_cod_deposit} calls the
        deposition API of TCOD and transfers the final CIF to the server
        for validation and deposition, if all checks are passed. The
        deposition step is optional, so that the final CIF can be simply
        exported as a local file on disk without deposition. This is
        useful for instance to manually inspect it before deposition,
        or if there is a need to share it privately.
\end{enumerate}

With \aiida{} installed, CIF generation can be achieved by running, 
on the command line, the command \texttt{verdi data structure 
export --format tcod PK}, where PK is the identifier of the 
structure to export. A large number of command line options exist to 
customise the behaviour, as explained in the \aiida{} documentation. 
CIF deposit can be achieved instead with the command \texttt{verdi 
data structure deposit}.

\section{Discussion}
As of March 2017, the number of records in TCOD has grown to more than
2\,200. As a proof of concept, over 170 theoretical structures have
been deposited to TCOD together with their provenance records using the
novel \aiida{}--TCOD interface presented here, consistuting 7\% of current
records in the TCOD. These depositions contain values of total energy in
addition to more than 4\,600 unique \aiida{} nodes, that are ready to be
automatically imported into user-side \aiida{} databases.

To ensure the completeness and semantic integrity of CIF files deposited
to TCOD, automatic checks are performed before accepting contributions. In
fact,  since CIF dictionaries contain formal descriptions of data items
and their values, they can be used for automatic validation of CIF
files~\cite{McMahon2006}. A number of tools for automatic CIF data
validation already exist, for example IUCr's
\script{checkCIF}~\cite{CheckCIF}, \emph{iotbx}'s
\script{cif.validate}~\cite{Gildea2011} and \emph{cod-tools}'s
\script{cif_validate}. The latter is developed by the COD and TCOD
development team and is the one used to validate files upon deposition. In
particular, as a part of this work, additional checks have been added to
\script{cif_cod_check}, the script (part of \emph{cod-tools}) responsible
for checking the semantic correctness of deposited data and for its
quality control: checks for the new data items added to the theoretical
dictionaries and verification that interrelated data items (data items for
coordinates; components of integration grid densities, shifts and residual
forces) are simultaneously present, when expected.

Furthermore, in Sec.~\ref{sub:encodings} we have introduced a number of
encodings for files that need to be included within CIF~1.1 files. While the
algorithms for decoding these files are known and available on the web, it
is cumbersome for a generic user to implement them in order to decode and
extract the files embedded in TCOD CIFs. To address this issue, we have
thus developed a program, \script{cif_tcod_tree}, to restore the full
directory tree used for the execution of the simulation in \aiida{},
stored in the TCOD CIF file as described in
Sec.~\ref{sec:provenancemodel}. After unpacking, the script further
fetches remote files that are not embedded in CIF using supplied URIs.
Finally, checksums are tested to ensure integrity of files. The program is
available for the end-users as part of \emph{cod-tools}, and can be used
to reproduce simulation results seamlessly.

Provided that software dependencies are met, the workflow can be
re-executed running each of its steps in the sequence specified in the
CIF file. Finally, output results can be easily compared with the original
values provided in the CIF. This makes it possible to run
unsupervised replication of deposited results to automatically assess the
validity of incoming data. We mention here, however, two aspects that need
to be kept in mind when implementing such a service. First, running again
the workflows could require a significant amount of computational time
(and, for some systems or properties, it is possible to run the workflows
only on large clusters). Moreover, particular care has to be taken to
prevent damage, accidental or deliberate, of the system replicating the
workflow, as well as illegal actions from the network. Runs should be
carried out only on isolated or limited systems (i.e., software jails,
virtual machines or Docker~\cite{Docker2016} images). For these reasons,
fully-automated workflow replication is not yet implemented in TCOD.
However, we foresee that validation of atomistic simulations could be
carried out ``on the cloud'' in a way very similar to continuous
integration services, even harnessing existing tools and infrastructures
such as Buildbot~\cite{Buildbot2016}, Jenkins~\cite{Jenkins2016} or
Travis-CI~\cite{Kalderimis2011}.

Finally, another important component that we have added to \aiida{}, 
as already discussed in Sec.~\ref{sec:inputdata}, is the set of 
importers from structure databases. These (and in particular those 
for COD and ICSD) have been already exploited as components of 
workflows for materials science high-throughput investigations. As 
an example, in~\cite{Mounet2016} the authors scanned both databases 
to discover, extract and screen 2D layered structures. 

We expect that these tools are going to be even more useful
for the computational community in the future, as they are distributed as
part of the open-source software \aiida{} and therefore freely available
to all researchers.

\section{Conclusions}
In this article we have shown the integration of the \aiida{} platform
(to automatically run and manage scientific workflows while keeping
full provenance of the computed data) and of TCOD (storing computed data
associated to crystal structures using an unambiguous ontology, 
within an open database to facilitate dissemination). 
Our integration makes it possible to obtain automatic \emph{a posteriori} 
tagging of crystal structures with metadata, like computed properties 
and their full provenance (codes adopted, inputs used in the computation, etc.).
We have first extended the TCOD CIF dictionaries for
atomistic simulations, \dic{cif_tcod} and \dic{cif_dft}, to include
provenance information.  We have then devised means to bypass the
intrinsic limitations of the CIF~1.1 file format adopted by TCOD. Moreover, we
have implemented provenance-aware importers into \aiida{} from a number of
external databases for crystal structures and pseudopotentials. The
main outcome of this work is the combination of all these efforts and the
implementation of a converter to automatically analyse the  data
provenance stored in \aiida{} after workflow execution, export the results
into a CIF file compliant with the TCOD dictionaries, and automatically
deposit it into TCOD. Additionally, we have developed a set of tools for
formal quality control and extraction of workflows from CIF files.

The general methodology described in this work does not have to be limited
to \aiida{} and TCOD, but may also be implemented in other frameworks and
databases. Our implementation proves that an automation
platform to manage simulations and automatically store 
the full provenance of computed datasets allows to add metadata 
at a later time, in a completely automated fashion.
Our integration of TCOD with \aiida{} constitutes
a fully-open platform implementing all four FAIR principles of
``Findability, Accessibility, Interoperability, Reusability'' for
scientific data management and stewardness~\cite{Wilkinson2016}, that is
furthermore fully interlinked with data generation. Indeed, our work
allows the deposition in an automated fashion of computational workflows
in an open database with permanent URIs and publicly-accessible
metadata/dictionaries (\emph{Findability}), that can be for instance
provided as supplementary material of computational papers. Data is
available over standard protocols like HTTP (\emph{Accessibility}) and,
thanks to the adoption of the established CIF format and its dictionaries,
both data and metadata are fully interlinked (\emph{Interoperability}).
Finally, data from the TCOD database together with its full provenance can
be easily retrieved and imported back into \aiida{} as input for further
calculations and analyses (\emph{Reusability}). We expect, therefore, that
in the future more researchers will adopt the methods and tools described
here to make the data public (as currently required by many funding
agencies) with minimal required effort.

\acknowledgements
The authors thank Philippe Schwaller for developing the ICSD importer
and Rickard Armiento for contributing theoretical data for deposition
to the TCOD.

This research was partially funded by the SCIEX grant No.~13.169, and by
the MARVEL National Centre of Competence in Research of the Swiss National
Science Foundation.

NMa and GP acknowledge partial support from the EU Centre of Excellence
``MaX - Materials Design at the Exascale'' (Horizon~2020 EINFRA-5,
Grant No.~676598).

\section*{Data and materials}
The datasets supporting the conclusions of this article are available
in the Crystallography Open Database,
\url{http://www.crystallography.net/cod} and the
Theoretical Crystallography Open Database,
\url{http://www.crystallography.net/tcod}.

\aiida{} is developed in Python for Unix-like operating systems. It
is released under MIT license and can be obtained free-of-charge from
\url{https://github.com/aiidateam/aiida_core}. This paper refers to
\aiida{}~\aiidav{}, a snapshot of which can be obtained from
\url{https://github.com/aiidateam/aiida_core/archive/v0.8.1.tar.gz}.

\emph{cod-tools} is developed in Perl and C for Unix-like operating
systems. It is released under GNU GPLv2 license and can be obtained
free-of-charge from \url{svn://www.crystallography.net/cod-tools}.
This paper refers to \emph{cod-tools}~v2.0, a snapshot of which can be
obtained from \url{http://www.crystallography.net/cod/archives/2017/software/cod-tools/cod-tools-2.0.tar.gz}.

\section*{Authors' contributions}
NMa and SG conceived the idea of the integration. AM developed the
required software and prepared the manuscript. Developments at the
\aiida{} and TCOD side were supervised by GP and SG, respectively.
NMo, AC and GP provided consultations concerning DFT and \qe{}, and
contributed 170 theoretical structures for deposition to the TCOD via
the novel interface. All authors contributed to the preparation of the
manuscript.

\bibliographystyle{bibstyle}
\bibliography{citations}

\end{document}